\documentstyle[12pt]{article}
\begin{document}
\begin{titlepage}

\title{On the Marriage of Chiral Perturbation Theory and Dispersion Relations}

\author{John F. Donoghue\\ [5mm]
Department of Physics and Astronomy\\
University of Massachusetts
Amherst, MA ~01003 ~U.S.A.}

\date{ }
\maketitle

\begin{abstract}
I describe the methodology for the use of dispersion relations in connection
with chiral perturbation theory.  The conditions for matching the two
formalisms are given at $O(E^2)$ and $O(E^4)$.  The two have several
complementary features, as well as some limitations, and these are
described by the use of examples, which include chiral sum rules related to
the Weinberg sum rules, form factors, and a more complicated reaction,
$\gamma \gamma \rightarrow \pi \pi$.

\end{abstract}
{\vfill UMHEP-417; hep-ph/9506205}
\end{titlepage}

Theoretical predictions are transformations from one set of known data to a
new set of data which we want to know.  In renormalizable field theories
one predicts many observables in terms of the results of a few experiments
measuring the fundamental constants $e$ and $m$.  In chiral perturbation
theory, measurements of a few low energy constants $(F_{\pi}, L_i)$,
which compactly summarize the effects of QCD, allow calculations of other
processes.  Dispersion relations make predictions in the same sense,
transforming measurable data in some process into predictions of the
physics elsewhere.

It is well known that in QCD one can make rigorous predictions at high
energy in a perturbative expansion in $\alpha_s$, although one also needs
structure functions which are not perturbatively calculable and which must
be measured.  It is becoming better known that rigorous results can
also be obtained at very low energy using chiral perturbation theory, with
calculations organized in an expansion in the energy.  However the
intermediate energy region is the most difficult.  This is traditionally the
realm of models, such as the quark model, Skyrme model etc. which, while
capturing some of the physics, suffer from a lack of control.  Dispersion
relations can sometimes be used to replace this model
dependence by experimental data.  In principle, dispersion
relations form a rigorous technique for this intermediate energy
region.  In practice, our knowledge of the input to dispersion relations is
often somewhat incomplete, so that some model dependence may enter, but
it can often be controlled or bounded to an acceptable level.  Done properly,
dispersive techniques will always enhance the reliability/range of chiral
calculations.

In this talk, I briefly describe chiral perturbation theory and dispersion
relations separately.  Although they have different starting points, it becomes
clear that the contents of chiral loops and of the dispersive integrals are
basically similar.  The remaining features of the two methods are
complementary and we can match the two descriptions in ways that are
mutually beneficial.  I describe how this matching occurs at different orders
in the energy expansion, and give examples of what is gained from doing
so.

\section{Chiral Perturbation Theory}

Chiral symmetry provides relations between amplitudes with different
numbers of zero energy pions.[1,2,3]  Corrections
to this limit can be given in an
expansion in the energy.  There exist various reduced matrix elements
which are not predicted by the symmetry and which therefore must be
measured, at least until reliable methods succeed in predicting them from
QCD.  According to the power counting theorem of Weinberg[3], at order
$E^2$ in the energy expansion one needs to consider tree level processes
and the only incalculable parameters are the pion mass, $m_{\pi}$, and its
decay constant, $F_{\pi}$.  At order $E^4$, one has a modest set of low
energy constants[2], $L_i, i=1 \ldots 10$, and in addition one must include
one loop diagrams.  At order $E^6$, one calculates to two loops and has a
formidable array of low energy constants.  It is unlikely that order $E^6$
calculations will be practical without the use of models to estimate these
parameters.

An example, which will also be useful later, is the pion electromagnetic
form factor.  At lowest order $O(E^2)$, one predicts simply that $f_{\pi}
(q^2) = 1$, while the tree level contribution at $O(E^4)$ involves the low
energy constant $L_9$ with a $q^2$ dependence, plus constant terms i.e.,

\begin{equation}
f^{(tree)}_{\pi} (q^2) = 1 + {2L_9 \over F^2_{\pi}} q^2 +{8m^2_{\pi}
 \over F^2_{\pi}} (2L_4 + L_5)
\end{equation}

\noindent Of the loop diagrams, Fig. 1b, c, that of Fig. 1b has no
$q^2$ dependence, contributing only a constant

\begin{equation}
\Delta f^{(1b)}_{\pi} (q^2) = {-5 m^2_{\pi} \over 48 \pi^2 F^2_{\pi}}
\left\{ {2 \over d-4} + \gamma - 1 - ln 4\pi + ln {m^2_{\pi} \over \mu^2}
\right\}
\end{equation}

\noindent Since we know that the pion form factor is absolutely normalized
to unity at $q^2 = 0$, we know that this constant must be canceled by
the wavefunction renormalization constant along with the diagram in
Fig. 1c, but the latter also
contains physics which is more interesting. Fig. 1c also contains important
dynamical information of the propgagation of the two pion state, including
the imaginary part of the amplitude due to on-shell intermediate states,
and the result involves a nontrivial function of $q^2$,

\begin{eqnarray}
\Delta f^{(1c)}_{\pi} (q^2) = {1 \over 16 \pi^2 F^2_{\pi}} \left\{ \left(
m^2_{\pi} - {1 \over 6} q^2 \right) \left[ {2 \over d-4} + \gamma
- 1 - ln 4 \pi + ln
{m^2_{\pi} \over \mu^2} \right] \right.\\ \nonumber
\left. + {1 \over 6} \left( q^2 - 4 m^2_{\pi} \right) H (q^2) - {1 \over
18}  q^2 \right\}
\end{eqnarray}

\noindent with

\begin{eqnarray}
H(q^2) &=& 2 + \beta \left[ ln \left( {1-\beta \over 1 + \beta} \right) + i \pi
\theta (q^2 - 4m^2_{\pi}) \right] \\ \nonumber
\beta &=& \sqrt{1-{4m^2_{\pi} \over q^2}}
\end{eqnarray}

\noindent Multiplying by the wavefunction renormalization constant

\begin{equation}
Z_{\pi} =1 -{8m^2_{\pi} \over F^2_{\pi}} (2L_4 + L_5)
 +{m^2_{\pi} \over 24\pi^2 F^2_{\pi} }
\left\{ {2 \over d-4} + \gamma - 1 - ln 4\pi + ln {m^2_{\pi} \over \mu^2}
\right\}
\end{equation}

\noindent and defining the renormalized value of $L_9$

\begin{equation}
L^r_9 = L_9 - {1 \over 192 \pi^2} \left[ {2 \over d-4} - ln 4 \pi + \gamma
-1 \right]
\end{equation}

\noindent we get the final result

\begin{equation}
f_{\pi} (q^2) = 1 + {2L^r_9 \over F^2_{\pi}} q^2 + {1 \over 96 \pi^2
F^2_{\pi}} \left[ (q^2 - 4m^2_{\pi}) H(q^2) - q^2 ln {m^2_{\pi} \over
\mu^2} - {q^2 \over 3} \right]
\end{equation}

What then is the content of chiral loops?  It is easy to state what are
{\em not}
important features of the loops.  First of all the high energy behavior of the
loop is not relevant, because we are using forms of the vertices which are
valid only at low energy. This ensures that the high energy portions of
the diagrams are not correct.  In a similar vein, the
divergences are not the important physics since they are part of the high
energy structure and do not correspond to the divergences of QCD
diagrams.  Fortunately, all high energy effects can be absorbed in a shift in
the low energy constants. This is true because the high energy portions must
obey the
symmetries of the theory and must be local effects when external particles
carry only low energy.  They are thus equivalent to a local term in
an effective Lagrangian. Likewise, diagrams such as Fig. 1b do not
have interesting physics because they just are universal constants.  In a
Feynman diagram approach, these are needed to enforce symmetry
properties.  But they play no dynamical role and if we had other ways to
enforce symmetry constraints, as we will in a dispersive approach, they
would not be needed.

The important loop physics comes from low energy intermediate states in
diagrams such as Fig. 1c.  This represents long range propagation and
cannot be mimicked by a shift of a parameter in a local chiral
Lagrangian.  Note the imaginary part which arises from this amplitude.
This represents the effects of unitarity coming from physical
intermediate states. Unitarity is satisfied order by order in the energy
expansion.  At one loop one uses the lowest order couplings in the vertex
and propagation without any rescattering in the intermediate  state.
Higher order loop
diagrams would allow for modification of the  vertices and for
rescattering in the intermediate state.

The outputs of chiral perturbation theory are relations between amplitudes,
order by order in the expansion in E, $m_q$.  At any given order, these
relations form low energy theorems of QCD.  However, in applying the
method in phenomenology, we often push it to regions where it is less
accurate than desirable.  In scattering amplitudes it is always tempting to use
chiral perturbation theory to describe reactions at higher energies until
eventually the result at a given order must
break down.  Similarly, in some calculations with kaons, the first known
corrections are large enough that we would like to know yet higher order
corrections.  Despite the beauty of the method, in phenomenological
applications the two main limitations are the fact that amplitudes are known
only to a limited order in the energy expansion and the proliferation of
unknown constants at order $E^6$ and higher.

\section{Dispersion Relations}

Scattering amplitudes and vertex functions will in general contain both real
and imaginary parts.[4,5]  The imaginary portions are due to the propagation of
on-shell intermediate states.  Causality implies certain properties of the
analytic structure of the amplitudes that it allows us to relate the real and
imaginary parts.  Such dispersion relations have the general form

\begin{equation}
Re f(s) = {1 \over \pi} P \int_{0}^{\infty} {ds' \over s' -
s}Im  f(s')
\end{equation}

\noindent With the identity

\begin{equation}
{1 \over x-x_0 - i \epsilon} = P {1 \over x-x_0} + i \pi \delta (x-x_0)
\end{equation}

\noindent one can write the full amplitude as an integral over its
imaginary part,

\begin{equation}
f(s) = {1 \over \pi} \int_{0}^{\infty} ds' {Im f(s') \over s' - s - i
\epsilon}
\end{equation}

Notice that the dispersive integral involves all $s'$.  In order to know
$f(s)$ at small $s$, we need to know $Im f(s')$ also at large $s'$.
We will see that subtractions can lessen the dependence on large
$s'$, but the integral still runs over all $s'$.  We in general
need to know the properties of on shell intermediate states.

All Feynman
diagrams share the required analytic structure and can be rewritten as
dispersion relations, perhaps with subtractions.  Therefore the content of
chiral loops can equally well be specified as a particular choice for $Im f(s
')$ in a dispersion integral.  When it is phrased this way, it is clear that
the content of chiral loop diagrams such as Fig. 1c and the content of a
dispersive integral are similar.  The chiral calculations uses a predicted
approximation to $Im f(s')$, while a properly performed dispersion
integral uses the real world data for $Im f(s')$. We will also see that the
chiral parameters $(L_i)$ play a similar role to the subtraction constants
in dispersion relations.

\section{Example:  The Weinberg sum rules and some relatives}

The simplest amplitudes are two point functions, and within QCD the
simplest of these are the particular combination of vector and axial vector
currents.

\begin{equation}
\pi^{\mu \nu}_V (q^2)-\pi^{\mu \nu}_A (q^2) \equiv i \int d^{4} x
e^{iq \cdot x} \langle 0 \mid T \left[ V^{\mu} (x) V^{\nu} (0)-A^{\mu} (x)
A^{\nu} (0) \right] \mid 0 \rangle
\end{equation}

\noindent This combination is analytic in the complex $q^2$ plane, except
for a pole at $q^2 = m^2_{\pi}$ and a cut for $q^2 > 4 m^2_{\pi}$.  The
vector current is conserved.  The axial current is conserved in the $m_q
\rightarrow 0$ limit, but with a Goldstone boson.  If we define scalar
function by

\begin{eqnarray}
\pi_V^{\mu \nu} (q^2) &=& (q^{\mu} q^{\nu} - g^{\mu \nu} q^2) \pi_V
(q^2) \\ \nonumber
\pi^{\mu \nu}_A (q^2) &=& (q^{\mu} q^{\nu} - g^{\mu \nu} q^2) \pi_A
(q^2) - q^{\mu} q^{\nu} \pi^{(0)}_A (q^2)
\end{eqnarray}

\noindent we can prove the dispersion relations

\begin{equation}
\pi_V (q^2) - \pi_A (q^2) = {F^2_{\pi} \over q^2} + \int^{\infty}_{4
m^2_{\pi}} ds' {\rho_V (s') - \rho_A (s') \over s'
- q^2 - i \epsilon}
\end{equation}

\noindent with the imaginary parts conventionally named via

\begin{equation}
\rho_{V/A} (s) = {1 \over \pi} Im
\pi_{V/A} (s)
\end{equation}

What is known theoretically about these amplitudes?  At low energy, chiral
perturbation theory predicts the form[2]

\begin{eqnarray}
\pi^{\mu \nu}_V (q^2) - \pi^{\mu \nu}_A (q) &=& \left[ {q_{\mu} q_{\nu}
\over q^2 - m^2_{\pi} + i \epsilon} - g_{\mu \nu} \right]  F^2_{\pi} \\
\nonumber
&+& (q_{\mu} q_{\nu} - g_{\mu \nu} q^2) \left[ {1 \over 48 \pi^2} \left( 1
- {4m^2_{\pi} \over q^2} \right) H (q^2) - 4L^r_{10} \right. \\ \nonumber
 &-& \left.{1 \over 48 \pi^2}  \left( ln {m^2_{\pi}
\over \mu^2} + {1 \over 3} \right) \right] \\ \nonumber
\rho_V (s) &=& {1 \over 48 \pi^2} \left[ 1 - {4 m^2_{\pi} \over s}
\right]^{3 \over 2} \theta (s - 4 m^2_{\pi}) + O(s) \\ \nonumber
\rho_A (s) &=& {s \over 96 (4 \pi F_{\pi})^2} + O(s^2)
\end{eqnarray}

\noindent Here $L^r_{10}$ is a low energy constant measured in radiative
pion decay, $\pi \rightarrow e \nu \gamma$.

At high energy, pertrubative QCD may analyse the two point function.  In
the chiral limit, $m_q = 0$, which will be used for the rest of this section,
the operator product expansion can be used to show that the difference
$\pi_V - \pi_A$ falls as $1 \over q^6$ and $\rho_V (s) - \rho_A (s) \sim
{1 \over s^3}$.  In terms of four quark operators, which are here evaluated
in the vacuum saturation approximation[6], one has

\begin{eqnarray}
\pi_V (q^2) - \pi_A (q^2) = {32 \pi \over 9} {\langle \sqrt{\alpha_s}
\bar{q} q \rangle^2_0 \over q^6} \left\{ 1 + {\alpha_s (q^2) \over 4 \pi}
\left[ {247 \over 12} + ln {\mu^2 \over -q^2} \right] \right\} \\ \nonumber
\rho_V (s) \rightarrow \rho_A (s) \rightarrow {1 \over 8 \pi^2} \left[ 1 +
{\alpha_s (s) \over \pi} \right] \; , \; s \rightarrow \infty \\ \nonumber
\rho_V (s) - \rho_A (s) \sim {8 \over 9} {\alpha_s \langle \sqrt{\alpha_s}
\bar{q} q \rangle^2_{\infty} \over s^3}
\end{eqnarray}

\noindent We see that $\pi_V - \pi_A$ and $\rho_V - \rho_A$ are very well
behaved at large $q^2, s$.

We can combine up this information to get a set of sum rules.  The
requirement that, as $q^2 \rightarrow \infty$, there is no ${1 \over q^2}$
term in the dispersion relation Eq.13 , requires

\begin{equation}
F^2_{\pi} = \int^{\infty}_0 ds ( \rho_V (s) - \rho_A (s))
\end{equation}

\noindent while the absense of ${1 \over q^4}$ implies

\begin{equation}
0 = \int^{\infty}_0 ds s ( \rho_V (s) - \rho_A (s))
\end{equation}

\noindent These are the Weinberg sum rules[7], the second of which is only
true in the $m_q \rightarrow 0$ limit.  At low energy, expansion of the
dispersion integral and chiral results in powers of $q^2$ imply[8,2]

\begin{equation}
-4 \bar{L}_{10} = \int^{\infty}_{4m^2_{\pi}} {ds \over s} ( \rho_V (s) -
\rho_A (s))
\end{equation}

\noindent with

\begin{eqnarray}
\bar{L}_{10} &=& L^r_{10} (\mu) + {1 \over 192 \pi^2} \left[ ln
{m^2_{\pi} \over \mu^2} + 1 \right] \\ \nonumber
&=& (-0.7 \pm 0.03) \times 10^{-2} \; (Expt: \pi \rightarrow e \nu \gamma)
\end{eqnarray}

\noindent Here I have given the sum rule for finite $m^2_{\pi}$ since there
is a behavior proportional to $ln m^2_{\pi}$ at the low energy
end of the integral.  These sum rules illustrate one of the uses of chiral
dispersion relations, which is the prediction/calculation of low energy
constants (here $F_{\pi}$ and $L_{10}$).

Another use of chiral dispersion relations is in extending the reach of
calculations and even opening up the possibility of entirely new types of
calculations.  Consider the Compton amplitude $\gamma \pi \rightarrow
\gamma \pi$.  In the soft pion limit, chiral symmetry relates this to the
vacuum polarization tensors

\begin{eqnarray}
&&\lim_{p \rightarrow 0} \langle \pi^+ (p) \mid T (V^{\mu} (x) V^{\nu}
(0)) \mid Ti^+ (p) \rangle \\ \nonumber
&=& - {1 \over F^2_{\pi}} \langle 0 \mid T (V^{\mu} (x) V^{\nu} (0) -
A^{\mu} (x) A^{\nu} (0) ) \mid 0 \rangle \\ \nonumber
&=& - {1 \over F^2_{\pi}} \left[ \pi^{\mu \nu}_V (x) - \pi^{\mu \nu}_A (x)
\right]
\end{eqnarray}

\noindent If one takes the Compton amplitude and ties together the two
electromagnetic currents with a photon propagator, one obtains the pion
electromagnetic
mass shift.[9]  Clearly the chiral representation, Eq. 15, would
be inadequate to calculate this, as the photon loop integral goes over all
values of $q^2$.  However, after some algebra plus the application
of the Weinberg sum rules, the dispersive representation allows one to
write this as

\begin{equation}
m^2_{\pi^+} - m^2_{\pi^0} = -{3e^2 \over 16 \pi^2 F^2_{\pi}}
\int^{\infty}_{0} ds s ln s \left[ \rho_V (s) - \rho_A (s) \right]
\end{equation}

\noindent which is an exact relation in the chiral limit. Note that here
chiral symmetry was used to relate different amplitudes in Eq. 23 and to
provide
low energy constraints , as in the Weinberg sum rules, while
dispersion relations were needed to provide a predictive
framework for the intermediate energy region.

In a similar way, one can calculate reliably a new weak nonleptonic matrix
element.[10]  Consider the hypothetical weak Hamiltonian

\begin{equation}
H_V = {g^2_2 \over 8} \int d^4 x i D^{\mu \nu}_F (x, M_w) T \left(
\bar{d} (x) \gamma_{\mu} u (x) \bar{u} (0) \gamma_{\nu} S (0) \right)
\end{equation}

\noindent Up to some KM factors, this would be the usual weak
Hamiltonian if the vector currents were replaced by $\gamma_{\mu} (1 +
\gamma_5)$.  In the chiral limit, we have another chiral sum rule

\begin{equation}
\langle \pi^- \mid H_V \mid K^- \rangle = {3i G_F \over 32 \pi^2 \sqrt{2}
F^2_{\pi}}A
\end{equation}

\noindent with

\begin{equation}
A = M^2_w \int^{\infty}_0 ds {s^2 ln (s/M^2_w) \over s - M^2_w + i
\epsilon} \left[ \rho_V (s) - \rho_A (s) \right]
\end{equation}

\noindent which is exact in the chiral limit.

Gene Golowich and I have recently provided a phenomenological analysis
of these sum rules.[11]  The physics of the spectral functions $\rho_{V,A}$ is
basically simple.  At intermediate energies they are measured in $\tau$ decay
and $e^+ e^-$ annihilation, and the largest features are the $\rho$ and
$a_1$ resonances, with very much smaller $4 \pi, 5 \pi$ etc. contributions.
At low energy this can be merged smoothly to chiral predictions and at high
energy $\rho_V - \rho_A$ vanishes rapidly and we matched the data to QCD
around $s = 5 GeV^2$.  There are some experimental uncertainties, but
these can in principle be reduced in the future.

The $L_{10}$ sum rule works well with very little uncertainty as it is
sensitive to the lowest energy contributions.  The Weinberg sum rules and
that for $\Delta m^2_{\pi}$ work within the experimental uncertainties.  We
have proceeded by imposing them exactly on our $\rho_V - \rho_A$, which
requires only minor adjustments within the allowed error bars.  That this is
possible is a nontrivial test of the theoretical framework.  Finally the weak
matrix element is predicted ($A = -0.062 \pm 0.017 GeV^2$).  This can
perhaps be compared with future lattice calculations.

\section{Subtractions}

Given a dispersion relation, one may also write a "subtracted" relation
for $(f (q^2) - f(0))/q^2$, i.e.,

\begin{equation}
{f(q^2) - f(0) \over q^2} = {1 \over \pi} \int {ds' \over s' - q^2
- i \epsilon} Im \left[ {f(s') - f(0) \over s'} \right]
\end{equation}

\noindent which, since $Im f(0) = 0$ is equivalent to

\begin{equation}
f(q^2) = f(0) + {q^2 \over \pi} \int {ds' \over s'} {Im f(s')
\over s' - q^2 - i \epsilon}
\end{equation}

\noindent This may be needed if $f(z) \neq 0$ at $\mid z \mid \rightarrow
\infty$, as a good behavior at infinity is required for the derivation of the
dispersion relation.  However, even if subtractions are not required, it may
still be desirable
to perform them.  Generally $Im f(s)$ is not well known at high
energy.  The subtracted dispersion integral weights lower energies more
heavily and lessens the influence of the high energy region.
The previous ignorance of the high energy effects of $Im f(s)$
is  reduced to a single number, the subtraction constant.  Further
subtractions  may be performed, with the introduction of further
subtraction constants.

The pion form factor obeys dispersion relations.  An unsubtracted form is

\begin{equation}
f_{\pi} (q^2) = {1 \over \pi} \int^{\infty}_{4 m^2_{\pi}} ds' {Im
f_{\pi} (s') \over s' - q^2 - i \epsilon}
\end{equation}

\noindent while with one subtraction the form is

\begin{equation}
f_{\pi} (q^2) = 1 + {q^2 \over \pi^2} \int^{\infty}_{4m^2_{\pi}} {ds'
\over s'} {Im f_{\pi} (s') \over s' - q^2 - i
\epsilon}
\end{equation}

\noindent Here the subtraction constant has been fixed to unity by the
normalization of the form factor.  A twice subtracted form is

\begin{equation}
f_{\pi} (q^2) = 1 + cq^2 + {q^4 \over \pi} \int^{\infty}_{4m^2_{\pi}} {ds'
\over s'^2} {Im f_{\pi} (s') \over s' - q^2 - i
\epsilon}
\end{equation}

\noindent where $c$ is presently unknown.

\section{Matching Conditions}

In dispersion relations involving subtraction constants we need a precise
identification of them.  Chiral perturbation theory provides an extensive
machinery for the analysis of the low energy behavior and can provide these
constants.[12]  The key is to reformulate chiral calculations as dispersion
relations, order by order.  As mentioned previously this is always possible
because the Feynman diagrams themselves satisfy dispersion relations.  An
important point is that the matching is different at order $E^2$ [13]
 and at order
$E^4$ [14,15].

At order $E^2$ one needs to reproduce only the tree level chiral results,
which do not involve imaginary parts.  Thus we only need to ensure that the
normalization at low energy is correct.  The dispersion integral will then
produce new effects at order $E^4$ which are equivalent to the prediction of
the low energy constants at order $E^4$, i.e., of the $L_i$.  This procedure
will be more sensitive to high energy effects because one will be using a
dispersion integral with at most one subtraction.

At order $E^4$ one knows more about the low energy structure so one can
use a dispersion relation with an extra subtraction.  The low energy
constants $L_i$ are no longer predicted, but are inputs to fix the subtraction
constants [The dispersion integral then produces new effects at order $E^6$
and higher].  To match at this order one must reproduce the one loop chiral
calculation.  Therefore the inputs to the dispersive integral must involve the
lowest order vertices, and will only have free propagations of the
intermediate state, i.e., the same inputs that go into the Feynman diagram
calculation.

As an example, let us first consider the pion form factor with matching at
order $E^4$.[14]  The one loop diagram, Fig. 1c, involves the $\pi \pi I = 1$
scattering amplitude, and the tree level $\pi \pi \rightarrow \gamma$ vertex,
so that

\begin{equation}
2(p_1 - p_2)_{\mu} Im f_{\pi} (s) = \int {d^3 p_1' d^3 p_2'
\over (2 \pi)^6 2E_1' 2E_2'} (2 \pi)^4 \delta^4 (s - p_1' -
p_2') \\ \nonumber
\langle \pi \pi \mid T \mid \pi \pi \rangle \langle
\pi \pi \mid  J_{\mu} \mid 0 \rangle
\end{equation}

\noindent or

\begin{equation}
Im f_{\pi} (s) = {1 \over 96 \pi F^2_{\pi}} {(s - 4m^2_{\pi})^{3 \over 2}
\over \sqrt{s}} \theta (s - 4 m^2_{\pi})
\end{equation}

\noindent We use the twice subtracted form, Eq. 30, and the dispersion
integral can be exactly done using

\begin{equation}
\int_{4m^2}^\infty  {ds \over s^2} \left( 1 - {4m^2 \over s}
\right)^{1 \over 2} \left( {a + b s
\over s -
q^2 - i \epsilon} \right) = {(a + b q^2) \over q^4}H (q^2) - {a \over
6m^2q^2}
\end{equation}

\noindent to give

\begin{equation}
f_{\pi} (q^2) = 1 + c q^2 + {1 \over 96 \pi^2 F^2_{\pi}} \left[ ( q^2 -
4m^2_{\pi}) H (q^2) + {2 \over 3} q^2 \right]
\end{equation}

\noindent Comparing this with the chiral calculation, Eq. 7, leads to the
identification of the subtraction constant

\begin{equation}
c = {2 L^{(r)}_9 \over F^2_{\pi}} - {1 \over 96 \pi^2 F^2_{\pi}} \left( ln
{m^2_{\pi} \over \mu^2} + 1 \right)
\end{equation}

\noindent If we now return to the full dispersive integral, we must be sure
that the input $Im f_{\pi} (s)$ agrees with the lowest order chiral result at
low energies, which of course it must in any case if chiral symmetry is a
valid description at low energy.  The use of the experimental $Im f_{\pi}
(s)$, thus constrained, then generates the full $f_{\pi} (q^2)$ at all $q^2$.
In principle, the only inaccuracy in this calculation is that we have given the
subtraction constant c by an expression which is exact only to order $E^4$.
There can be corrections to this by extra factors of $m^2_{\pi}$ or
$m^2_{\pi} ln m^2_{\pi}$.

Let us also briefly consider the same quantity matched at $O(E^2)$, using
Eq. 29.  Now the only matching is the simple constraint $f_{\pi} (0) = 1$,
and the effect of the dispersive integral starts at $q^2$.  This leads to a
prediction of the low energy constant

\begin{equation}
2L^r_9 + {1 \over 96 \pi^2} \left( ln {m^2_{\pi} \over \mu^2} + 1 \right) =
F^2_{\pi} \int^{\infty}_{4 m^2_{\pi}} {ds' \over s'^2} Im
f_{\pi} (s')
\end{equation}

\noindent Note that the lowest order form of $Im f_{\pi} (s)$ cannot be
inserted in the once subtracted dispersion integral, as the result diverges.
The lowest order form for $Im f_{\pi} (s)$ is not valid at high energies, but
the twice subtracted integral used above was not sensitive to this.
The use of the real data for $Im f_{\pi}(s')$ leads to a succesful prediction
of $L^r_9$ in terms of the mass of the rho meson.

{}From these examples, we can see clearly the dynamical content of
dispersion relations.  If minimally subtracted (i.e., just barely convergent)
this includes the effects of low, moderate and high energy intermediate
states.  If oversubtracted (i.e., more than is required by the high energy
behavior), high energy effects are damped and we retain the effects of low
and moderate energy propagation, and can remain consistent with the chiral
constraints while extending the calculation to higher $q^2$.

Thus in the best of all worlds (full data on $Im f(s)$, many related reactions)
the two techniques form a powerful combination which allows rigorous
results at all energies.  Chiral perturbation theory provides the subtraction
constants from symmetry relations and dispersion relations allows the
extrapolation to higher energy.

In real world phenomenology, we often have incomplete information.  In
addition, we may want/need to predict $Im f$ also.

\section{The Elastic Approximation and the Omnes Problem}

Consider some two particle amplitude $f(s)$ of a given isospin and angular
momentum which is analytic in complex $s$ plane except for a cut above
two particle threshold $s_0 = 4m^2$.  The inelastic thresholds are somewhat
higher, for example $s_{inel} = 16 m^2$.  In the elastic region, Watson's
theorem tells us that the phase of the amplitude is that of the corresponding
two particle scattering amplitude

\begin{equation}
f(s) = e^{i \delta (s)} \mid f(s) \mid
\end{equation}

\noindent In practice inelasticities do not play a significant role in low
energy pion physics up to 1 GeV ($K \bar{K}$ threshold), and one often
assumes an approximation of keeping only the elastic channel.  While
probably reasonable, it is important to realize that the elastic approximation
relies on more than just Watson's theorem and produces more than just the
phase of the amplitude.

The Omnes problem[16] is the mathematical exercise of finding functions which
are analytic except for a cut $4m^2 < s < \infty$, which are real when
$s$ is real and less than $4m^2$ and for which $f(s) e^{-i \delta (s)}$ is real
when $s$ is real and greater than $4m^2$.  The solution is given by

\begin{eqnarray}
f(s) &=& P(s) D^{-1} (s) \\ \nonumber
D^{-1} (s) & \equiv & exp \left\{ {s \over \pi} \int^{\infty}_{4m^2} {dt
\over t} {\delta (t) \over t - s - i \epsilon} \right\}
\end{eqnarray}

\noindent as long as

\begin{equation}
\lim_{s \rightarrow \infty} \delta (s) = finite \; ;
\lim_{s \rightarrow \infty}
{\mid f (s) \mid \over s} \rightarrow 0
\end{equation}

\noindent In the above $P(s)$ is a polynomical in $s$, and $D^{-1} (s)$ is
called the Omnes function.

Note that this is not exactly the right problem for QCD.  The assumption
that $f(s) e^{-i \delta (s)}$ is real above the cut implies that the reaction
is elastic at {\em all} energies.  Once inelastic channels open up, the
quantity $f(s)e^{-i\delta (s)}$ rapidly deviates from being real.
In QCD, once one is above 1
GeV, the inelastic channels open rapidly and become quite numerous, leading
to perturbative QCD behavior at precociously low energies.  It is not
known how to provide a general solution to the QCD type problem (although the
form of the solution to the two channel problem is also known), nor is it known
how much of an effect the multiple inelasticies of QCD have on the Omnes
function.

\section{Example:  Matching at order $E^2$}

One of the earliest examples of the utility of merging chiral perturbation
theory and dispersion relations came in the analysis of the possible decay of
a light Higgs to
two pions.[13]  This is no longer of phenomenological interest,
but the technique developed illustrates the basic methodology.  Among other
couplings, one is faced with the matrix of the energy momentum tensor,

\begin{equation}
\langle \pi^i (p) \pi^j (p')) \mid \theta_{\mu}^{\mu} \mid 0 \rangle =
\theta_{\pi} (s) \delta^{ij}
\end{equation}

\noindent At lowest order, chiral perturbation theory tells us that

\begin{equation}
\theta_{\pi} (s) = s + 2m^2_{\pi}
\end{equation}

\noindent while at order $E^4$ the form is

\begin{equation}
\theta_{\pi} (s) = (s + 2m^2_{\pi}) \left\{ 1 + \phi (s) \right\} + b_{\theta}
s^2 + O(m^4_{\pi}, sm^2_{\pi} , E^6)
\end{equation}

\noindent where $b_{\theta}$ is a combination of the low energy constants
$L_{11}, L_{12}, L_{13}$ that occur when one analyses the energy
momentum tensor[17] and $\phi (s)$ is a known loop function.  The trouble is
that we have no phenomenology which measures $b_{\theta}$.  However,
$\theta_{\pi} (s)$ can be shown to satisfy a dispersion relation, and the
elastic region
only involves $\pi \pi$ scattering in the $I = 0, J = 0$ channel.
Therefore, in the elastic approximation, we can reproduce both the right
chiral properties and satisfy the Omnes solution by choosing

\begin{equation}
\theta_{\pi} (s) = (2 m^2_{\pi} + s) D^{-1} (s)
\end{equation}

\noindent [In practice, a two channel solution was found, involving $K
\bar{K}$ states above 1 GeV.  However, for our example, let us neglect
$K \bar{K}$.]  The Omnes function was constructed from $\pi \pi$ data by
Gasser.[18]

The output of this representation is a prediction for $\theta_{\pi} (s)$ at
higher energies than is possible with chiral perturbation theory alone.  One
of the by products is a prediction of the low energy constant.

\begin{equation}
b_{\theta} = 2.7 GeV^{-2}
\end{equation}

\noindent In this particular channel, the effects of rescattering are
significant.  For example at $\sqrt{s} = 0.5 GeV$, the lowest order formula
gives

\begin{equation}
\theta_{\pi} (s) = 0.29
\end{equation}

\noindent while at one loop, $O(E^4)$ with the above value of $b_{\theta}$
one finds

\begin{equation}
\theta_{\pi} (s) = 0.46 + 0.13_i
\end{equation}

\noindent and the full dispersive amplitude is

\begin{equation}
\theta_{\pi} (s) = 0.40 + 0.31_i
\end{equation}

\noindent This large effect is typical of the $I = 0, J = 0 ~\pi \pi$
amplitude,
which gets large very quickly, so much so that the lowest order chiral
prediction for it violates unitarity around 600 MeV.  When the $I = 0, J = 0
{}~\pi \pi$ scattering channel is important in a calculation, a dispersive
treatment could be useful.  It is also typical that one can obtain good
agreement for the magnitude $\mid \theta_{\pi} (s) \mid$ with a suitably
chosen low energy constant at $O(E^4)$, but that the imaginary part will be
too small at one loop when compared to the full answer.  This is because the
one loop imaginary part corresponds to using the lowest order amplitudes
in the unitarity relation.  This
is not a big problem, because it could be corrected by hand as the second
order amplitudes are known and we could use these to determine the second order
imaginary part.  However, the dispersive treatment does this automatically,
as well as including yet higher orders.

\section{Example:  Matching at Order $E^4$}

The reaction $\gamma \gamma \rightarrow \pi^+ \pi^-$ and $\gamma
\gamma \rightarrow \pi^0 \pi^0$  are of interest in the development of chiral
theory because $\gamma \gamma \rightarrow \pi^0 \pi^0$ first arises as a
pure loop effect as there are not tree level contributions at $O(E^2)$ or
$O(E^4)$.  For these
reactions, we have both a one-loop [19]and two loop [20]chiral
analysis as well as
dispersive treatments[21,15] and experimental data.  This makes
these reactions excellent
illustrations of chiral techniques and of the ties with
dispersion relations.

The $\gamma \gamma \rightarrow \pi \pi$ matrix elements can be
decomposed into isospin amplitudes

\begin{eqnarray}
f^{+-}(s) &=& {1 \over 3} [2f_0(s) + f_2(s)] \\ \nonumber
f^{\infty}(s) &=& {2 \over 3} [f_0(s) - f_2(s)]
\end{eqnarray}

\noindent The dominant partial waves at low energy are the S waves and
these are predicted in a one loop chiral analysis to be

\begin{eqnarray}
f^{chiral}_I (s) &=& {1-\beta^2 \over 2 \beta} ln \left( {1 + \beta \over 1 -
\beta} \right) - {(1 - \beta^2) \over 4 \pi} t^{CA}_I (s) ln^2 {\beta + 1 \over
\beta - 1} \\ \nonumber
&& - {1 \over \pi} t^{CA}_I (s) + {2 \over F^2_{\pi}} \left( L^r_9 +
L^r_{10} \right) s
\end{eqnarray}

\noindent where

\begin{equation}
\beta = \sqrt{1 - {4m^2_{\pi} \over s}}
\end{equation}

\noindent and $t^{CA}_I (s)$ are the lowest order $\pi \pi$ scattering
amplitudes

\begin{equation}
t^{CA}_0 = {2s-m_{\pi} \over 32 \pi F^2_{\pi}} \; ; t^{CA}_2 = -{(s -
sm^2_{\pi}) \over 32 \pi F^2_{\pi}}
\end{equation}

The dispersion relation has been derived by Morgan and Pennington[21], in
terms of an amplitude $p_I (s)$ which has the same left-hand singularity
structure as $f_I(s)$ but which is real for $s > 0$.  Then $[f_I(s) - p_I(s)]
D_I(s)$ satisfyies a twice subtracted dispersion relation and we have

\begin{equation}
f_I(s) = D^{-1}_I (s) \left[ (C_I + d_I s) + p_I (s) D_I (s) - {s^2 \over \pi}
\int^{\infty}_{4m^2_{\pi}} {ds' \over s'^2} {p_I (s')
Im D_I (s') \over s' - s - i \epsilon} \right]
\end{equation}

\noindent with two subtaction constants per channel $c_I, d_I$.  As a
prelude to the matching we note that Low's theorem requires that $f_I(s)$
be the Born scattering amplitude at low energies.  Therefore

\begin{equation}
p_I(s) = f^{Born}_I(s) + O(s) = {1- \beta^2 \over 2 \beta} ln \left( {1 +
\beta \over 1 - \beta} \right) + O(s)
\end{equation}

\noindent This is the $O(E^2)$ result.  To proceed to order $E^4$ we note
that the leading piece of $Im D_I(s)$ is also known, i.e.,

\begin{equation}
Im D_I(s) = - \beta t^{CA}_I (s)
\end{equation}

\noindent as this is the lowest order $\pi \pi$ scattering amplitude.  Using
this, the dispersive integral can be done exactly, leading to

\begin{eqnarray}
f_I(s) &=& D^{-1} (s) \left[ c_I + s \left( d_I - {t^{CA}_I (0)
\over 12 \pi  m^2_{\pi}} \right) + D_I(s) {1 - \beta^2 \over 2 \beta}
ln \left( {1 + \beta  \over 1 - \beta} \right) \right. \\ \nonumber
&-&\left. {1 \over 4 \pi } (1 - \beta^2) t^{CA}_I (s) ln^2 \left( {\beta
+ 1 \over \beta - 1} \right) \right] + \ldots
\end{eqnarray}

\noindent A comparison of this with the $O(E^4)$ chiral results then
indicates that this procedure has reproduced all of the one loop results, as
long as we choose the subtraction constants as[15]

\begin{equation}
c_I = 0 \; ; d_I = {2 \over F^2_{\pi}} \left( L^r_9 + L^r_{10} \right) +
{t^{CA}_I (0) \over 12 \pi m^2_{\pi}}
\end{equation}

\noindent Again we see that the dynamical content of the one loop chiral
calculation is also contained in the dispersive treatment when the imaginary
part is taken to be the lowest order scattering amplitude.  However, chiral
symmetry also predicts the subtraction constants, which in this case are
known from measurements in radiative pion decay.

Having identified the subtraction constants one can add the ingredients to
complete the calculation.  The most important at threshold is the use of the
real world $D^{-1}_I (s)$[18].  This change alone produces a significant effect
in the amplitude
even near threshold in the neutral case.
The second step is a better determination of
$p_I(s)$ which includes the $O(E^4)$ chiral corrections to it as well as the
$\rho, \omega, A1$ poles which are known (from $\rho \rightarrow \pi
\gamma$ etc. data) to occur in the Compton amplitude.
Figure 2 shows the data for the
reaction $\gamma \gamma \rightarrow \pi^0 \pi^0$ along with the one-loop
chiral prediction (dashed line) and the modification obtained by the dispersive
treatment (solid line). The one-loop
chiral result is of the right rough size, its slope is low at
threshold and it grows unphysically at high energy.
Near threshold  the difference in the two calculations
comes almost exclusivly from the rescattering corrections generated through
the dispersion relation. The change is sizeable even at low energy, since the
rescattering in in the $I=0, J=0$ channel.  .  The Omnes function
alone has brought the threshold region into better agreement with the data.
It has also  tamed the
high energy growth. The final result
(with no free parameters) matches the data very well, and also
gives the charged channel correctly.

Belucci, Gasser and Sainio[20] have performed the enormously difficult two
loop calculation.  [In fact, technically they employ dispersive methods to do
portions of this.]  At two loop order, new low energy constants appear,
which are not measured in any other process.  Therefore the authors have to
step outside of pure chiral perturbation theory in order to model these
constants, using vector meson dominence.  Much like the dispersive work
described above, these constants play little role in the threshold region, but
are important for the shape of the amplitude for moderate energy.  It is
very interesting that their results look very similar to the
dispersive treatment described above.

Both of these methods have potential limitations.  In principle, the only
limitation of the dispersive treatment is the fact that it can miss $O(E^6)$
terms in the subtraction constants $c_I, d_I$.  These would be corrections
to results given above by factors of $m^2_{\pi}$ or $m^2_{\pi} ln
m^2_{\pi}$.  In practice we also need to model the higher order terms in
$p_I(s)$.  As for
the two loop chiral result, its only limitation in principle is
the fact that it misses higher order dependnce in the energy.  By
construction it is valid to order $E^6$, but does not contain
higher order $s$ dependence, and
so would be expected to fall apart soon after the $E^6$ dependence became
important.  In practice, this approach also needs to do some modeling in
order to estimate the unknown low energy constants.  The fact that the
results agree so well with each other and with the data indicates that these
limitations are not very important at these energies.  Both capture the
important physics, and do so in a reasonably controlled fashion.  There is of
course a significant practical advantage to the dispersive approach--it is far
easier!

\section{Summary}

We have seen how dispersion relations can add power to chiral perturbation
theory.  At
its best it uses more physics input.  It can match all chiral effects
to whichever order that they are known, and can be used to replace the
modelling of unknown physics by using data instead of models.  However
there are some limitations, coming both from incomplete data and from the
fact that we can only determine the subtraction constants to a given order in
the energy expansion.

The technology for combining these techniques is now developed.  This
involves first knowing the chiral analysis of the amplitude to a given order
in the energy expansion.  One also needs a dispersion relation for the
amplitude in question.  The number of subtractions is determined partially
by the high energy behavior of the amplitude, but the use of more subtractions
than are required can help in the matching with the chiral result.  The
matching occurs order by order in the energy expansion.  When it can be
done, it is preferable to perform the matching at $O(E^4)$
because the resulting dispersive treatment
is less sensitive to what happens at high energy since a twice subtracted
dispersion relation can be used.  Finally, the real world data has to be found
to use in the dispersive integral.  Often, the use of the elastic approximation
is made for this, allowing the use of known $\pi \pi$ scattering data.

The output of these efforts can be several.  Most commonly, these
techniques are used to extend the range and accuracy of the chiral
calculations, by getting around the limitation of the energy expansion.  The
method can be used to predict unknown chiral coefficients, as was shown
for the case of $L_{10}$.  We can use these techniques to remove or reduce
the model dependence of some result.  Finally, dispersive techniques allow
us to perform completely new types of calculations, such as the hadronic
matrix elements of Section IV.

There is work going on which is pushing the frontier of what can be done
using these techniques.  More difficult reactions, such as $K \rightarrow \pi
\gamma \gamma$ or $K \rightarrow \pi e^+ e^-$, require more subtle analyses.
Probably more important, the use of dispersive and chiral techniques in
hadronic matrix elements can likely be pushed further.  Dispersion relations
and chiral perturbation theory bring different strengths to their union, and
the marriage, although not without an occasional flaw, has been friutful.

{\bf Acknowledgements}.  I would like to thank Juerg Gasser, Barry
Holstein and Heiri Leutwyler for sharing their insights into dispersion
relations and chiral symmetry.  In addition, I thank C.H. Lee and D.P. Min
for their kind hospitality at this conference.

{\bf References}.\\
1) J. F. Donoghue, E. Golowich and B. R. Holstein, {\it Dynamics of the
Standard Model}, (Cambridge University Press, Cambridge, 1992).\\
2) J. Gasser and H. Leutwyler, Nucl. Phys. {\bf B250}, 465(1985).\\
3) S. Weinberg, Physica {\bf A96}, 327 (1970).\\
4) R. Kronig, J. Op. Soc. Am. {\bf 12}, 547 (1926); H. A. Kramers,
Atti Cong. Int. Fisici Como (1927).\\
5) G. Barton, {\it Dispersion Techniques in Field Theory}, (Benjamin, NY,
1965).\\
6) M. A. Shifman, A. I. Vainshtein, and V. I. Zakharov, Nucl. Phys. {\bf B147},
385 (1979).\\
L. V. Lanin, V. P. Spiridonov, and K. G. Chetyrkin, Sov. J. Nucl. Phys.
{\bf 44}, 892 (1986).\\
7) S. Weinberg, Phys. Rev. Lett. {\bf 17}, 616 (1966).\\
8) T. Das, V. Mathur, and S. Okubo, Phys. Rev. Lett. {\bf 19}, 859 (1967).\\
9) T. Das, G. S. Guralnik, V. S. Mathur, F. E. Low, and J. E. Young,
Phys. Rev. Lett. {\bf 18}, 759 (1967).\\
10) J. F. Donoghue and E. Golowich, Phys. Lett. {\bf 315}, 406 (1993).\\
11) J. F. Donoghue and E. Golowich, Phys. Rev. {\bf D49}, 1513 (1994).\\
12) As far as I know, the first attempt to use dispersion relations to
improve the predictions of chiral perturbation theory occured in Ref. 13.
The techniques at order $E^4$ have been illustrated in Ref 14,15.\\
13) J. F. Donoghue, J. Gasser and H. Leutwyler, Nucl. Phys. {\bf B343},
341(1990).\\
14) T. N. Truong, Phys. Rev. Lett. {\bf 61},2526 (1988).\\
15) J. F. Donoghue and B. R. Holstein, Phys. Rev. {\bf D48}, 137 (1993).\\
16) R. Omnes, Nuovo Cim. {\bf 8}, 1244(1958).\\
17) J. F. Donoghue and H. Leutwyler, Z. Phys. {\bf C52},343(1991).\\
19) J. Bijnens and F. Cornet, Nucl. Phys. {\bf B296}, 557 (1988).\\
J. F. Donoghue, B. R. Holstein and Y. C. Lin, Phys. Rev. {\bf D37}, 2423
(1988).\\
20) S. Bellucci, J. Gasser and M. Sainio, Nucl. Phys. {\bf B423},80{1994).\\
21) D. Morgan and M. R. Pennington, Phys. Lett. {\bf B272}, 134 (1991).\\
\end{document}